\documentclass[aps,prl,twocolumn,showpacs,groupedaddress,floatfix]{revtex4}
\usepackage{times,amsmath,amsfonts,amsthm}
\usepackage{graphicx}
\usepackage{dcolumn}
\usepackage{bm}
\usepackage{amssymb}

\begin{document}

%% &&&&&&&&&&&&&&&&&&&&&&&&&&&&&&&&&&&&&&&&&&&&&&&&&&&&&&&&&&&&&&&&&&&&&&&&&&&&&&&&
%% &&&&&&&&&&&&&&&&&&&&&&&&&&&&&&&&&&&&&&&&&&&&&&&&&&&&&&&&&&&&&&&&&&&&&&&&&&&&&&&&
%%
%%                           Title
%%
%% &&&&&&&&&&&&&&&&&&&&&&&&&&&&&&&&&&&&&&&&&&&&&&&&&&&&&&&&&&&&&&&&&&&&&&&&&&&&&&&&
%% &&&&&&&&&&&&&&&&&&&&&&&&&&&&&&&&&&&&&&&&&&&&&&&&&&&&&&&&&&&&&&&&&&&&&&&&&&&&&&&&

\title{Positive Cross Correlations in a Normal-Conducting
Fermionic Beam Splitter}

\author{S.\,Oberholzer}
\email{stefan.oberholzer@unibas.ch}
%\homepage{www.unibas.ch/phys-meso}
\author{E.\,Bieri}
\author{C.\,Sch\"onenberger}
\affiliation{Institut f\"ur Physik, University of Basel,
Klingelbergstr.\,82, CH-4056 Basel, Switzerland}

\author{M.\,Giovannini}
\author{J.\,Faist}
\affiliation{Institut de Physique, Universit\'{e} de Neuch\^atel,
Rue A.\,L.\,Breguet\,1, 2000 Neuch\^atel, Switzerland}

\date{2 February 2006}

%% &&&&&&&&&&&&&&&&&&&&&&&&&&&&&&&&&&&&&&&&&&&&&&&&&&&&&&&&&&&&&&&&&&&&&&&&&&&&&&&&
%% &&&&&&&&&&&&&&&&&&&&&&&&&&&&&&&&&&&&&&&&&&&&&&&&&&&&&&&&&&&&&&&&&&&&&&&&&&&&&&&&
%%
%%                           Abstract
%%
%% &&&&&&&&&&&&&&&&&&&&&&&&&&&&&&&&&&&&&&&&&&&&&&&&&&&&&&&&&&&&&&&&&&&&&&&&&&&&&&&&
%% &&&&&&&&&&&&&&&&&&&&&&&&&&&&&&&&&&&&&&&&&&&&&&&&&&&&&&&&&&&&&&&&&&&&&&&&&&&&&&&&

\begin{abstract}
We investigate a beam splitter experiment implemented in a normal
conducting fermionic electron gas in the quantum Hall regime. The
cross-correlations between the current fluctuations in the two
exit leads of the three terminal device are found to be negative,
zero or even \emph{positive} depending on the scattering mechanism
within the device. Reversal of the cross-correlations sign occurs
due to interaction between different edge-states and does not
reflect the statistics of the fermionic particles which
`antibunch'.
\end{abstract}


%\keywords{shot-noise,quantum-Hall effect}
\maketitle

%% &&&&&&&&&&&&&&&&&&&&&&&&&&&&&&&&&&&&&&&&&&&&&&&&&&&&&&&&&&&&&&&&&&&&&&&&&&&&&&&&
%% &&&&&&&&&&&&&&&&&&&&&&&&&&&&&&&&&&&&&&&&&&&&&&&&&&&&&&&&&&&&&&&&&&&&&&&&&&&&&&&&
%%
%%                           Introduction
%%
%% &&&&&&&&&&&&&&&&&&&&&&&&&&&&&&&&&&&&&&&&&&&&&&&&&&&&&&&&&&&&&&&&&&&&&&&&&&&&&&&&
%% &&&&&&&&&&&&&&&&&&&&&&&&&&&&&&&&&&&&&&&&&&&&&&&&&&&&&&&&&&&&&&&&&&&&&&&&&&&&&&&&

In mesoscopic physics, the investigation of the conductance of
small electronic devices is widely used to obtain their transport
properties. Additionally to such time-averaged measurements, the
temporal fluctuations (noise) in the current caused by the
granularity of charge and diffraction of the wave-function provide
us with  important supplementary information about electronic
transport \cite{BlanterReview}. Nonequilibrium noise has been
widely explored to determine, for example, the effective charge of
carriers \cite{Glattli,Picciotto,Kozhevnikov,Jehl} or to study the
transmission properties of quantum coherent devices such as
quantum point contacts \cite{Reznikov,Kumar}, diffusive wires
\cite{eindrittel} or chaotic cavities \cite{einviertel}.
Universally, the statistical correlations due to the Pauli
exclusion principle are responsible for the negative
current-current correlations between different leads in
\emph{multi-terminal} devices \cite{Buettiker1992,Martin1992}.
Such negative correlations have been observed in distinct
experiments \cite{Science1999_1,Science1999_2,PHE2000,Kiesel2002}.
It has also been shown that for a diluted electronic stream
obeying classical statistics the negative correlations vanish
\cite{PHE2000}.

In contrast, a positive cross-correlation (i.e.\,`bunching') is
predicted to occur in devices with `non-normalconducting' contacts
\cite{Buettiker2003} like hybrid-structures which use a
superconductor as current-injector: two entangled electrons,
forming a Cooper in the superconductor, are simultaneously emitted
into different exit leads giving rise to a positive correlation
\cite{Anantram,Martin,Torres,Gramespacher,Recher,Borlin,Samuelsson}.
Alternatively, devices with ferromagnetic contacts can show
positive cross-correlations due to `opposite spin-bunching'
\cite{SauretPRL2004} or dynamical \mbox{spin-blockade
\cite{CottetPRL2004}.}

\begin{figure}[h]
\includegraphics[width=77mm]{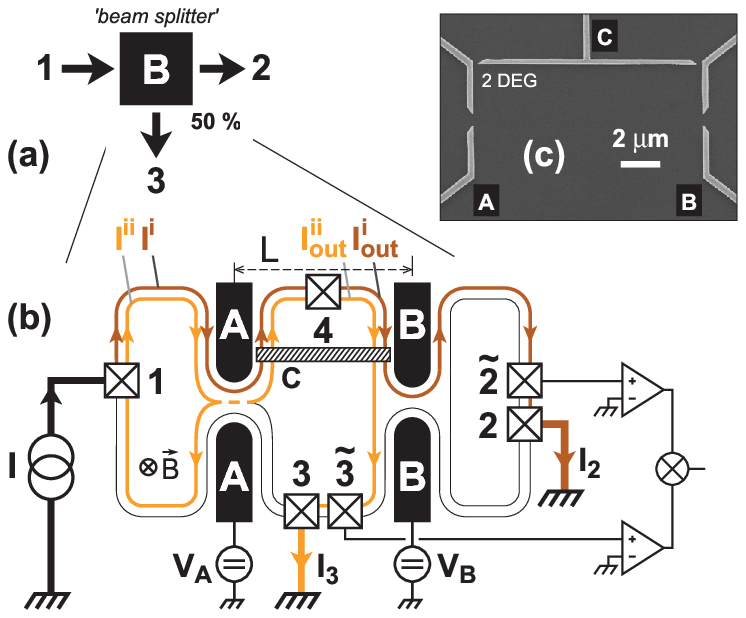}\vspace{-4pt}
\caption[fig1]{(a,b) Implementation of the beam splitter in the
quantum Hall regime with two edge-states ($i$,\,$ii$)
($B$\,=\,$1.6$\,Tesla). Equilibration between them occurs along
the path from $A$ to $B$ when the current flows into an additional
voltage probe 4. (c) SEM image: QPC's are formed by
 split gates on top of a two-dimensional electron
gas.}\label{fig1}\vspace{-14pt}
\end{figure}

In this article we are interested in a discussion by Texier and
B\"uttiker \cite{TexierPRB} about the occurrence of positive
cross-correlations in a purely \emph{normal-conducting fermionic}
device. As the authors show, the effect is due to current
redistribution among different conducting states. We consider this
idea here experimentally in a beam-splitter configuration
(Fig.\,\ref{fig1}(a)) where a current $I$ injected at contact 1 is
split into two equal parts exiting into contacts 2 and 3. Our main
result is the observation of positive cross-correlations between
the two exit contacts for a particular implementation of the
beam-splitter according to Ref.\,\cite{TexierPRB}. Although
predicted by various theoretical works,  such positive
cross-correlations have not been seen before in mesoscopic
devices. We further conclude from our experimental observation
that a positive correlation in fermionic systems can be
interpreted as a sign of entanglement only if effects as the one
shown here can be ruled out.

Fig.\,\ref{fig1}(b) gives an `inside view' of the physical
implementation of the beam splitter configuration used to study
the cross-correlations sign reversal from negative to positive: A
two-dimensional electron gas (2DEG) is exposed to a perpendicular
magnetic field so that the current flows in edge-states along the
border of the device \cite{Halperin1982,Buettiker1988}.
Edge-states provide natural fermionic `beams' which are, thanks to
their chirality, easily split by a quantum point contact (QPC)
into a transmitted and a reflected part \cite{Buettiker1992}. The
two (tunable) QPC's in series play different roles: the first one
($A$) introduces noise within the edge-state(s) while the second
one ($B$) acts as a beam splitter that exits the edge-states or
parts of them into different contacts. In former experiments, only
\emph{one single} spin-degenerated edge-state was populated
\cite{PHE2000}. It was shown that the correlations are always
negative as expected for a single stream of fermionic particles
showing `antibunching' behavior \cite{PHE2000,TexierPRB}.
\noindent

In the following, we consider the case that exactly the
last two spin-degenerated Landau levels are fully occupied. Partitioning at %QPC
$A$ with the transmission probability $T_{A}^{ii}$ gives rise to
current fluctuations $\Delta I^{ii}$ in the second
\mbox{edge-state ($ii$).} Their power spectral density is
$\langle(\Delta I^{ii})^2\rangle_{\omega}$\,$=$\,$2\,G_0\,
T_{A}^{ii}(1$\,$-$\,$T_{A}^{ii})\,\mu_1$ with $\mu_1$ the
electrochemical potential of contact 1 and $G_0$\,=\,$2e^2/h$
\cite{Khlus1987,Lesovik1989,Buettiker1990}. The first edge-state
remains noiseless because it is transmitted at $A$ with unit
probability ($T_{A}^{i}$\,$\equiv$\,1). Inter-edge-state
equilibration introduced via an extra voltage probe 4
redistributes the current fluctuations $\Delta I^{ii}_{in}$ in the
current $I^{ii}_{in}$ incident to the mixing contact 4 between the
two outgoing edge-states $I^{i}_{out}$ and $I^{ii}_{out}$: $\Delta
I^{i}_{out}$\,$=$\,$\Delta I^{ii}_{out}$\,$=$\,$\Delta I^{ii}/2$.
Finally, the 'beam splitter' $B$ separates the two edge-states
into two different contacts 2 and 3. Since the current
fluctuations in both edge-states originate from the same
scattering process at $A$ the cross-correlations are expected to
be \emph{positive}. Their power spectral density $\langle\Delta
I_{2}\Delta I_{3}\rangle_{\omega}$\,=$\langle\Delta
I^{i}_{out}\Delta I^{ii}_{out}\rangle_{\omega}$ divided by the
Poissonian value $2e|I|$ \mbox{equals \cite{TexierPRB}:}
\begin{equation}\label{positive}
    \frac{\langle\Delta I_{2}\Delta I_{3}\rangle_{\omega}}{2e|I|}
    = \frac{\langle(\Delta I^{ii})^2\rangle_{\omega}}{8e|I|}=
    +\frac{1}{4}\frac{T_{A}^{ii}(1-T_{A}^{ii})}{1+T_{A}^{ii}}.
\end{equation}
Here, $I$\,=\,$G_0\,(1$\,$+$\,$T_{A}^{ii})\,\mu_1/e$ describes the
total current injected at contact 1.
%%
%%
%% &&&&&&&&&&&&&&&&&&&&&&&&&&&&&&&&&&&&&&&&&&&&&&&&&&&&&&&&&&&&&&&&&&&&&&&&&&&&&&&&
%% &&&&&&&&&&&&&&&&&&&&&&&&&&&&&&&&&&&&&&&&&&&&&&&&&&&&&&&&&&&&&&&&&&&&&&&&&&&&&&&&
%%
%%                           Experimental
%%
%% &&&&&&&&&&&&&&&&&&&&&&&&&&&&&&&&&&&&&&&&&&&&&&&&&&&&&&&&&&&&&&&&&&&&&&&&&&&&&&&&
%% &&&&&&&&&&&&&&&&&&&&&&&&&&&&&&&&&&&&&&&&&&&&&&&&&&&&&&&&&&&&&&&&&&&&&&&&&&&&&&&&
%%
%%

\begin{figure}[t]
\includegraphics[width=82mm]{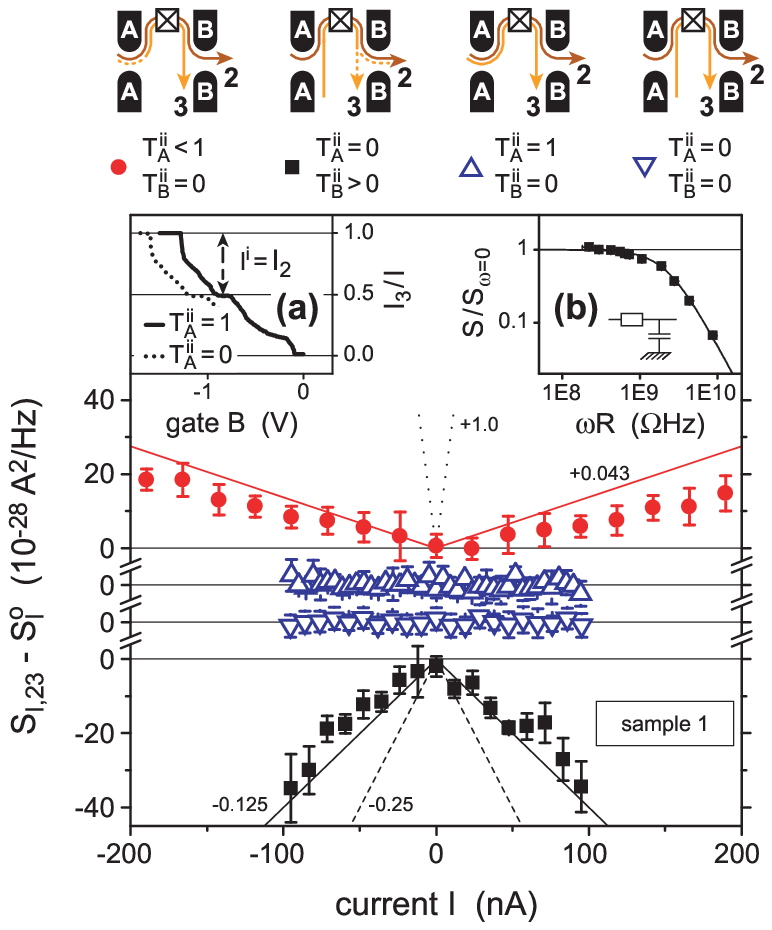}\vspace{-8pt}
\caption[fig2]{Current fluctuations due to scattering at $A$ are
redistributed among the two edge-states so that positive
cross-correlations are observed. Partial scattering at the second
point contact $B$ reveals the fermionic nature of the edge-states
and yields a negative correlation. The correlations are zero in
case that no partial scattering occurs at the QPC's. The data
($\triangledown$) are measured on sample 2. (a) Reflected current
at $3$ with gate $C$ closed. (b) RC-damping of the voltage
noise.}\label{fig2}\vspace{-14pt}
\end{figure}

Experimentally, the device illustrated in Fig.\,\ref{fig1}(b) is
implemented in a standard
\mbox{GaAs/}\-\mbox{Al$_{0.3}$Ga$_{0.7}$As}-he\-te\-ro\-structure.
The QPC's $A$ and $B$ are defined by metallic split-gates on top
of the 2DEG, which forms 60\,nm below the surface
(Fig.\,\ref{fig1}(c)). Two samples with different path lengths $L$
(200 and 14\,$\mu$m) between the two QPC's have been measured. The
solid curve in Fig.\,\ref{fig2}(a) shows the normalized reflected
current $I_3/I$ as function of the voltage applied to gate $B$
with gate $A$ open. It is given by
$I_3/I$\,=\,$1$\,-\,$(T_B^i$\,+\,$T_B^{ii})/2$. For
$I_3/I$\,$<$\,$0.5$ we obtain the transmission $T_B^{ii}$ by
measuring $I_3$ ($T_B^{i}$\,$\equiv$\,1). The \mbox{transmission
$T_A^{ii}$} is determined similarly.
%%
%% &&&&&&&&&&&&&&&&&&&&&&&&&&&&&&&&&&&&&&&&&&&&&&&&&&&&&&&&&&&&&&&&&&&&&&&&&&&&&&&&
%%                          Figure 2
%% &&&&&&&&&&&&&&&&&&&&&&&&&&&&&&&&&&&&&&&&&&&&&&&&&&&&&&&&&&&&&&&&&&&&&&&&&&&&&&&&
%%
\noindent
%%
%% &&&&&&&&&&&&&&&&&&&&&&&&&&&&&&&&&&&&&&&&&&&&&&&&&&&&&&&&&&&&&&&&&&&&&&&&&&&&&&&&

In order to detect the current-current cross-correlations between
contact 2 and 3 the time dependent currents $I_{\alpha}(t)$
($\alpha$\,=\,2,\,3) are converted to voltage signals
$V_{\tilde{\alpha}}(t)$ by two series resistors
$R_{\tilde{\alpha}\alpha}$\,$=$\,$h/4e^2$\,$+$\,$R_{0,\alpha}$
implemented by means of additional ohmic contacts \~{2} and \~{3}.
$R_{0,\alpha}$ denotes the contact resistances of the ohmic
contacts, which is of the order \mbox{0.5\,-\,3\,k$\Omega$.} The
voltage fluctuations $\Delta V_{\tilde{\alpha}}(t)=\Delta
I_{\alpha}(t)R_{\tilde{\alpha}\alpha}$ are measured by two
low-noise amplifiers and fed into a spectrum analyzer which
calculates the power spectral density. The $RC$-damping of the
voltage noise due to the finite capacitance of the measurement
lines (Fig.~\ref{fig2}(b)) and the offset-noise $S_0$ due to the
amplifiers are obtained from a calibration measurement of the
Nyquist noise $4k_B\theta R$ as function of the bath temperature
$\theta$ for a given resistance $R$. The noise measurements are
performed in a frequency range of 20 to 70\,kHz with typical
bandwiths of 5\,kHz. The measurement frequencies as well as the
current bias are chosen such that contributions from $1/f$-noise
are negligible. All measurements were performed in a
$^3$He-cryostat with a base temperature of 290\,mK. %%
%%
%% &&&&&&&&&&&&&&&&&&&&&&&&&&&&&&&&&&&&&&&&&&&&&&&&&&&&&&&&&&&&&&&&&&&&&&&&&&&&&&&&
%% &&&&&&&&&&&&&&&&&&&&&&&&&&&&&&&&&&&&&&&&&&&&&&&&&&&&&&&&&&&&&&&&&&&&&&&&&&&&&&&&
%%
%%                           Results & Discussion
%%
%% &&&&&&&&&&&&&&&&&&&&&&&&&&&&&&&&&&&&&&&&&&&&&&&&&&&&&&&&&&&&&&&&&&&&&&&&&&&&&&&&
%% &&&&&&&&&&&&&&&&&&&&&&&&&&&&&&&&&&&&&&&&&&&&&&&&&&&&&&&&&&&&&&&&&&&&&&&&&&&&&&&&
%%
%%

Fig.\,\ref{fig2} gives the cross-correlations
\mbox{$S_{I,23}=\langle\Delta I_{2}\Delta I_{3}\rangle_{\omega}$}
 measured on sample 1 for different configurations of gate $A$ and
$B$ and with gate $C$ open. In a first measurement $T^{ii}_A$
equals $\simeq$\,0.5 and the beam splitter $B$ is adjusted such
that the second edge-state is totally reflected
($T^{ii}_B$\,=\,0), which corresponds to the configuration shown
in Fig.\,\ref{fig1}(b). For these parameters we indeed observe a
\emph{positive} cross-correlation (red full-circles). The solid
line is the maximal positive cross-correlation given by
Eq.\,(\ref{positive}). For comparison the Poissonian-noise
$S_0$\,$=$\,$2e|I|$ is given as dotted line. The total offset
$S^0_I$ equals \mbox{$3.13\cdot 10^{-27}$\,A$^2$/Hz.} The
current-noise of the amplifiers gives an offset
$S^{0}_{I,\theta=0}$ of $3.91$\,$\cdot$\,$10^{-27}$\,A$^2$/Hz,
which we obtain from several temperature calibrations. From these
two values the thermal correlations between contact 2 and 3 can be
calculated
$S_{I,23}$($I$=0$)$\,=\,$S^0_I$\,$-$\,$S^0_{I,\theta=0}$\,=\,\mbox{$-7.9$\,$\cdot$\,$
10^{-28}$\,A$^2$s}, which turn out to be negative. Thermal
correlations are always negative
\cite{Buettiker1992,Science1999_1}. They are not related to the
statistics of the charge carriers but occur due to charge
conservation. The measured value is in reasonable agreement with
the theoretical prediction of $-k_B\theta\,G_0
(3-T^2)$\,=\,$-8.8\cdot 10^{-28}$\,A$^2$s for $T^{ii}_B$\,=\,$0$,
$T^{ii}_A$\,=\,$T$\,=\,$0.5$ and \mbox{$\theta$\,=\,$290$\,mK
\cite{TexierPRB}.}

For transparencies $T^{ii}_B$\,$>$\,0 the second edge-state is
only partially reflected at the `beam splitter'. Consequently, the
statistical properties of the electrons in the fermionic `beam'
become apparent and the cross-correlations change sign from
positive to negative. The solid line indicates the `full
antibunching' of $-2e|I|/8$ for $T^{ii}_B$\,=\,0.5 and
$T^{ii}_A$\,=\,1 or 0. Naturally, for
$T^{ii}_A,\,T^{ii}_B$\,$\in$\,$\{0,1\}$ the cross-correlations are
zero indicating that inelastic scattering between two edge-states
alone does not introduce any noise in the system.
%%
%% &&&&&&&&&&&&&&&&&&&&&&&&&&&&&&&&&&&&&&&&&&&&&&&&&&&&&&&&&&&&&&&&&&&&&&&&&&&&&&&&
%%                          Figure 3
%% &&&&&&&&&&&&&&&&&&&&&&&&&&&&&&&&&&&&&&&&&&&&&&&&&&&&&&&&&&&&&&&&&&&&&&&&&&&&&&&&
%%
\begin{figure}[t]
\includegraphics[width=82mm]{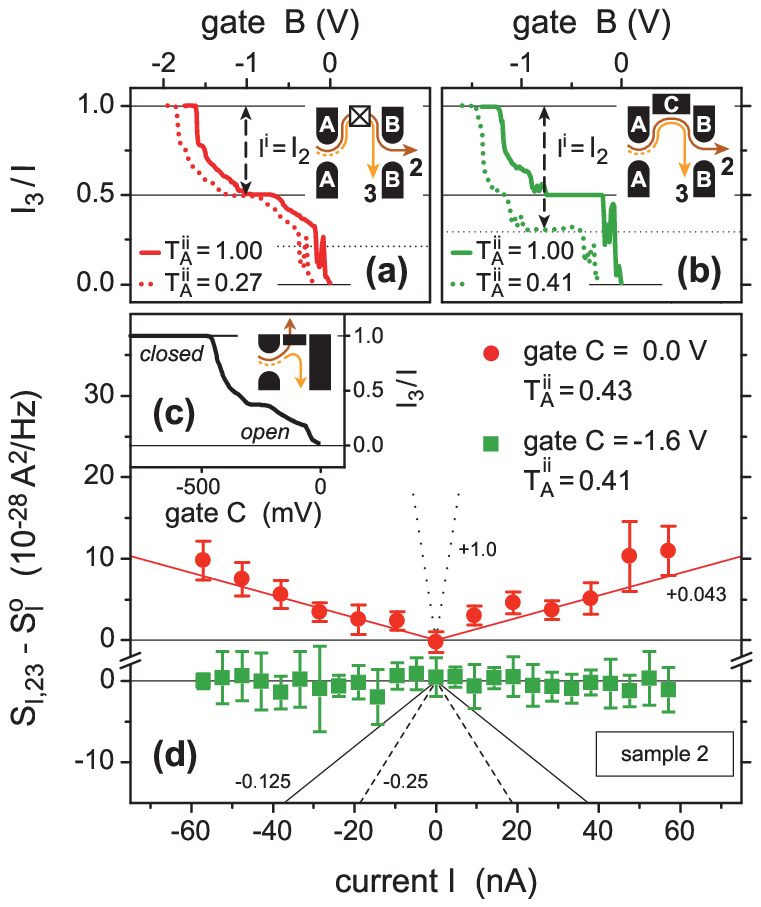}\vspace{-8pt}
\caption[fig3]{(a,\,b) Normalized current $I_3/I$ for different
transmissions $T^{ii}_A$. The curves are shifted in $V$ for
clarity. (a) The edge-states equilibrate between $A$ and $B$
within a floating voltage probe. (b) Gate $C$ closed: no
equilibration occurs and the currents carried by the two
edge-states are unequal if the second edge-state is only partially
transmitted at $A$. (c) $I_3/I$ vs. gate $C$ with gate $B$ closed
and $T^{ii}_A$\,$\simeq$\,$0.4$. Contact 4 is on ground. (d) A
positive cross-correlation is observed with gate $C$ open which
disappears for gate $C$ closed}\label{fig3}\vspace{-14pt}
\end{figure}\noindent
%%
%% &&&&&&&&&&&&&&&&&&&&&&&&&&&&&&&&&&&&&&&&&&&&&&&&&&&&&&&&&&&&&&&&&&&&&&&&&&&&&&&&

Next we will discuss what happens if there is no equilibration
present. Fig.\,\ref{fig2}(a) gives the reflected current at
contact 3 with gate $C$ \emph{closed} for $T^{ii}_A$\,=\,0 and 1.
At the observed plateau, the second edge-state is totally
reflected. Thus, it's height yields a direct measure for the
amount of current carried by the edge-states \cite{Alphenaar1990}.
Although for $T^{ii}_A$\,=\,0, no current is carried by the second
edge-state the plateau at $I_3/I$\,=\,0.5 indicates current
redistribution along the path $L$\,=\,$\overline{AB}$ from $A$ to
$B$ of length $\simeq$\,200\,$\mu$m. This is in agreement with
detailed studies on equilibration lengths in the quantum Hall
regime \cite{vanWees1989,Alphenaar1990}. In order to avoid
equilibration, we have to consider another device where the path
length $L$ (=14\,$\mu$m) of the two QPC's is much shorter.
Fig.\,\ref{fig3}(a,\,b) show the reflected current at contact 3
measured on this second device \mbox{(sample 2).} In (a) the
edge-states equilibrate in contact 4, like in the first
experiment, and the current is redistributed among the two
edge-states so that the plateau does not change in height. In
Fig.\,\ref{fig3}(b) however, gate $C$ is closed and the current
$I^{ii}$ carried by the second edge-state now depends on the
transmission $T^{ii}_A$. The dotted lines correspond to
$I^{ii}/I$\,=\,$I_3/I$\,=\,$T^{ii}_A/(1$\,+\,$T^{ii}_A)$. The
plateau which appears at the height of the dotted line in
Fig.\,\ref{fig3}(b) thus proves that the two edge-states do not
equilibrate. This should be noticed in the noise, too.
Fig.\,\ref{fig3}(d) presents cross-correlation measurements with
$T^{ii}_B$\,$=$\,$0$ and $T^{ii}_A$\,$\simeq$\,$0.42$. With
equilibration in contact 4 (gate $C$\,=\,0.0\,V) the correlations
are positive (full circles) and in good agreement with the maximal
positive correlation. If gate $C$ is closed the first edge-state
remains noiseless ($\Delta I^{i}$\,=\,$\Delta I_2$\,=\,0) and the
correlator $\langle\Delta I^{i}\Delta I^{ii}\rangle_{\omega}
$\,$=$\,$\langle\Delta I_{2}\Delta I_{3}\rangle_{\omega}$ vanishes
(squares). We thus have a `knob' which allows us to turn the
positive correlations on and off.

\begin{figure}[t]
\includegraphics[width=82mm]{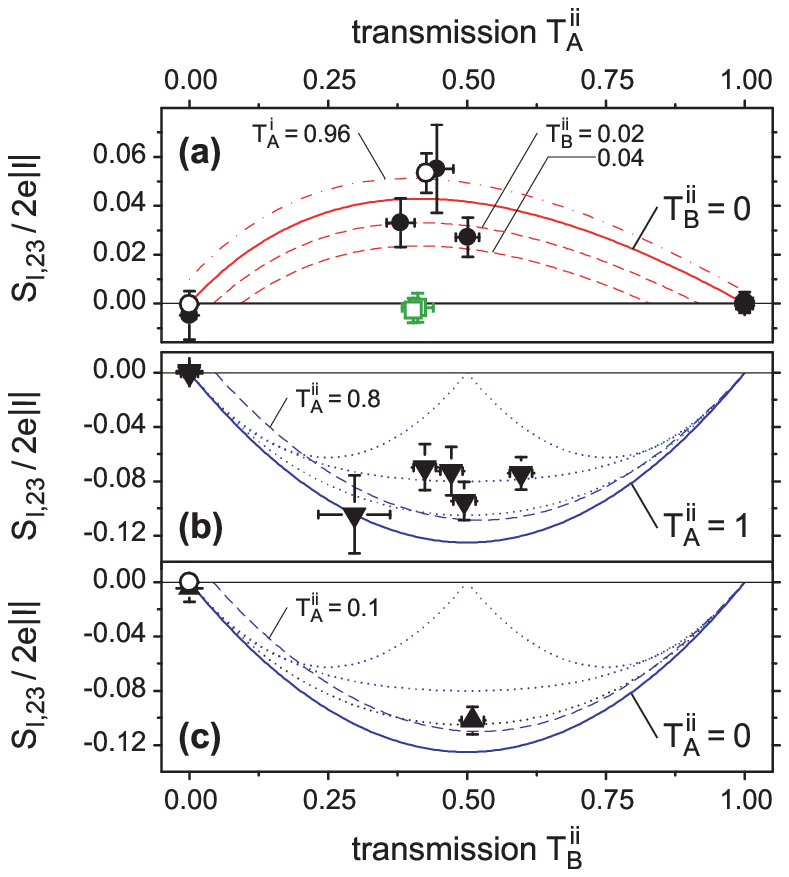}\vspace{-8pt}
\caption[fig4]{Cross-correlation for various
$T^{ii}_A$,\,$T^{ii}_{B}$ extracted from noise measurements on two
different samples (filled symbols: $ \overline{AB}
\simeq$\,200\,$\mu$m, open symbols:
$\overline{AB}\simeq$\,14\,$\mu$m. (a) Positive cross-correlations
occur for $T^{ii}_B=0$. (b,\,c) For $T^{ii}_A$\,$=$\,$0$ or 1 the
correlations are always negative.}\label{fig4}\vspace{-14pt}
\end{figure}\noindent

In Fig.\,\ref{fig4} we compare the cross-correlations
$S_{23}/2e|I|$ measured on the two different samples for various
parameters $T^{ii}_A$ and $T^{ii}_B$ with theoretical calculations
from Ref.\,\cite{TexierPRB}. At zero temperature the correlations
between contact 2 and 3 are described by:
\begin{eqnarray}\label{S23}
  \frac{S_{I,23}}{2e|I|} &=& -\frac{\sum T_B^n(1-T_B^n)}{2}\,+\nonumber\\
  &&\frac{(\sum T_B^n)(2-\sum T_B^n)}{4}
    \,\frac{\sum T_A^n(1-T_A^n)}{\sum T_A^n},
\end{eqnarray}
where $n$\,=\,$i$,\,$ii$ denotes the index of the two edge-states.
The first term in Eq.\,(\ref{S23}) describes the negative
correlations due to partitioning of the edge-states at $B$ whereas
the second term gives a positive contribution due to
inter-edge-state equilibration. With
$T_A^{i}$\,=\,$T_B^{i}$\,$\equiv$\,1, Eq.\,(\ref{S23}) yields a
maximal possible positive cross-correlation of
\mbox{$(3/4$\,-\,$1/\sqrt{2})$\,$\simeq$\,$0.043$\,$\cdot$\,$S_0$}
that occurs for $T_A^{ii}$\,=\,$\sqrt{2}$\,$-$\,$1$ with
$T_B^{ii}$\,=\,$0$. In Fig.\,\ref{fig4}(a) the measured positive
correlations are somewhat smaller for sample 1 (black circles).
For sample 2 with a smaller path length $L$ between the QPC's the
data points (open circles) are rather close to the expected value.
Although the QPC $B$ is adjusted to a plateau with high precision
the second edge-state $(ii)$ might not be reflected perfectly.
Already a tiny transmission $T_B^{ii}$ of 2\,\% reduces the
maximal positive cross-correlation by 23\,\%, illustrated by one
of the dashed curves in Fig.\,\ref{fig4}(a). The positive
correlations completely disappear for $T^{ii}_B$\,$>$\,$9$\,\%.
The high sensitivity to any changes from $T_B^{ii}$\,$=$\,$0$ thus
might explain the deviations from the solid curve. The open
squares in Fig.\,\ref{fig4}(a) are the results from sample 2 where
gate $C$ is closed so that the state-mixing voltage probe 4 is
disconnected. Cross-correlations larger than
$0.043$\,$\cdot$\,$S_0$ could theoretically occur due to
additional scattering of the first edge-state at $A$. The
dashed-dotted curve gives an example for $T_A^i$\,=\,0.96 instead
of 1 that would yield a maximal positive correlation of
$0.051\cdot S_0$.

Fig.\,\ref{fig4}(b,\,c) summerize the negative correlations
obtained for $T^{ii}_A$\,$=$\,$1$ and $0$, respectively. The data
points do not exactly agree with the expected values according
Eq.\,(\ref{S23}) (solid curves). The dashed curves denote the
changes that would occur due to additional scattering at the first
\mbox{QPC $A$}, yielding a small positive contribution to the
negative \mbox{correlations \cite{PHE2000}}. However, the
transmission at $A$ equals 0 or 1 (open gate) with quite high
precision ($|\Delta T_{A}^{ii}|$\,$\leq $\,$0.03$) and we think
that the deviations observed here are related to non-equal
transmissions of the two spin-polarized parts in the second
edge-state. The dotted lines in Fig.\,\ref{fig4}(b,c) are the
negative correlations for 20, 40 and 100\,\% unequal transmission
(from bottom to top). For one spin-polarized edge-state totally
transmitted and the other totally reflected the correlations would
be zero for $\langle T_B^{ii}\rangle $\,=\,0.5. From the data we
estimate that the differences between the two transmissions are in
the order of \mbox{20\,-\,40\,\% of $T_B^{ii}$.}

In conclusion we have observed positive cross-correlations in a
multi-terminal electronic device. This positive correlations occur
due to interactions between different current carrying states
inside the device and can be switched on and off by means of an
external gate voltage, which controls the interaction inside the
device.

The authors thank M.\,B\"uttiker, C.\,Hoffmann, and M.\,Calame for
valuable comments. This work has been supported by the Swiss
National Science Foundation and the NCCR on Nanoscience. MG and JF
thank the NCCR on quantum photonics.

%%
%%
%% &&&&&&&&&&&&&&&&&&&&&&&&&&&&&&&&&&&&&&&&&&&&&&&&&&&&&&&&&&&&&&&&&&&&&&&&&&&&&&&&
%% &&&&&&&&&&&&&&&&&&&&&&&&&&&&&&&&&&&&&&&&&&&&&&&&&&&&&&&&&&&&&&&&&&&&&&&&&&&&&&&&
%%
%%                           References
%%
%% &&&&&&&&&&&&&&&&&&&&&&&&&&&&&&&&&&&&&&&&&&&&&&&&&&&&&&&&&&&&&&&&&&&&&&&&&&&&&&&&
%% &&&&&&&&&&&&&&&&&&&&&&&&&&&&&&&&&&&&&&&&&&&&&&&&&&&&&&&&&&&&&&&&&&&&&&&&&&&&&&&&
%%
%%

\end{document}